# Experimental violation of the Leggett-Garg inequality using the polarization of classical light

Wenlei Zhang ,*,† Ravi K. Saripalli * Jacob M. Leamer , Ryan T. Glasser,[1,‡] and Denys I. Bondar §
*Department of Physics and Engineering Physics, Tulane University, New Orleans, Louisiana 70118, USA*

(Received 3 February 2021; accepted 20 October 2021; published 28 October 2021)

In contrast to Bell's inequalities, which test the correlations between multiple spatially separated systems, the Leggett-Garg inequalities test the temporal correlations between measurements of a single system. We experimentally demonstrate the violation of the Leggett-Garg inequality in a classical optical system using only the polarization degree of freedom of a laser beam. Our results show maximal violations of the Leggett-Garg inequality.

## I. INTRODUCTION

In their seminal paper [1], Leggett and Garg formulated a test for violations of the postulates of macrorealism based on the temporal correlations between measurements of a single system. The original postulates are stated as (A1) macroscopic realism per se—a macroscopic system with two or more macroscopically distinct states available to it will at all times be in one or the other of these states—and (A2) noninvasive measurability—it is possible, in principle, to determine the state of the system with arbitrarily small pertubation on its subsequent dynamics. For a macrorealistic system and some dichotomic observable $Q$ (with realizations $\pm 1$), the correlation function $C_{ij} \equiv \langle Q(t_i)Q(t_j)\rangle$, where $Q(t_i)$ is the measurement of $Q$ at time $t_i$, must satisfy a class of inequalities, which are now referred to as the Leggett-Garg inequalities (LGIs). In the simplest scenario, the LGI is given by [2]

$$C_{21} + C_{32} - C_{31} \leqslant 1. \tag{1}$$

Leggett and Garg originally proposed experimental tests using superconducting quantum interference devices [1]. Since then, violations of the LGIs have been observed in experiments based on superconducting transmon qubits [3,4], nuclear spin qubits [5–10], and light-matter interaction [11]. Experimental violations of the LGIs have also been demonstrated in optical systems, most of which measure properties of single photons [12–16], whereas Ref. [17] has experimentally demonstrated the violation of an LGI with the polarization of classical light. However, the authors of Ref. [17] utilize a particular initial state, namely, the eigenstate of $Q(t_1) = +1$, to test a modified version of the LGI shown in Eq. (1). They employ a noninvasive measurement at $t_2$, which consists of a CNOT operation with the orbital angular momentum as the controlling bit and polarization as the controlled bit. Effectively, the portions of the classical light beam with horizontal and vertical polarizations are split into two different paths. In a recent theoretical proposal [18], the authors considered the possibility of violating an LGI with coherent states in a random walk setup which utilizes the path degree of freedom. All the previous examples of violations of the LGIs, with the exception of [17,18], are based on measurements of microscopic systems. While the study of violations of the LGIs in microscopic systems is interesting in its own right, we feel this is not sufficient evidence to refute macrorealism in the eye of a stubborn macrorealist.

In this work, we present an experimental violation of the LGI in Eq. (1) in a classical optical system using only the polarization degree of freedom of a laser beam. Unlike Ref. [17], we demonstrated the violation of the LGI in Eq. (1) using different initial states of polarization. In the quantum optical sense, a classical state of light is defined as a convex (i.e., incoherent) sum of coherent states, and a laser beam is the closest approximation to a coherent state experimentally. The power of the laser beam used in our experiment is $\sim$100 mW, which corresponds to $\sim 10^{17}$ photons per second. A system with such a large number of particles should be considered a macroscopic system. Throughout this paper, the classical state of polarization (or simply "state") of the laser beam is described using the coherency matrix generalization of the Jones calculus [19–22] and quantities with a hat symbol ($\hat{}$) denote matrices within this context. Our results show maximal violations of the LGI in Eq. (1).

## II. THE JONES CALCULUS AND THE COHERENCY MATRIX

The Jones calculus [19] is commonly used to describe the interaction between polarized light and linear optical components. In the Jones calculus, the state of polarized light is represented by a Jones vector, and linear optical components are represented by Jones matrices. For a monochromatic transverse wave traveling in the $+z$ direction, at a particular point in the transverse profile, the Jones vector is given by a column vector,

$$|\psi\rangle = \begin{pmatrix} E_x(t) \\ E_y(t) \end{pmatrix}, \tag{2}$$

where $E_x(t)$ and $E_y(t)$ are the complex amplitudes of the electric field in the $x$ and $y$ directions, respectively. After passing through a linear optical component represented by a $2 \times 2$

*These authors contributed equally to this work.
†wzhang13@tulane.edu
‡rglasser@tulane.edu
§dbondar@tulane.edu

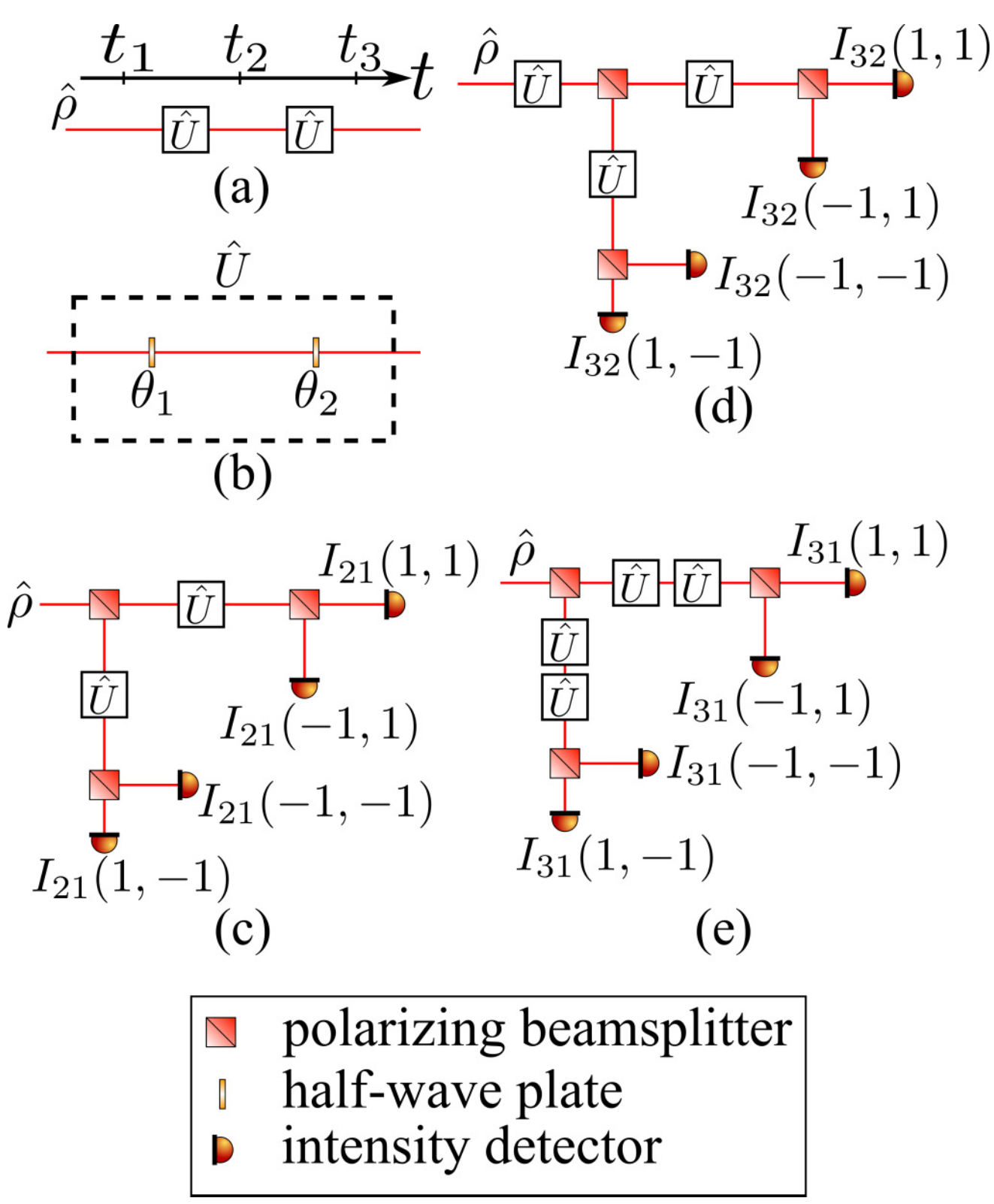

FIG. 1. (a) Time evolution of the coherency matrix $\hat{\rho}$. (b) Realization of a unitary operation $\hat{U}$ on the polarization matrix using two half-wave plates. Experimental setups for measuring (c) $C_{21}$, (d) $C_{32}$, and (e) $C_{31}$.

Jones matrix $\hat{J}$, the initial Jones vector $|\psi\rangle$ is transformed into $\hat{J}|\psi\rangle$.

While the Jones vector represents polarized light, it can be generalized to describe partially polarized and unpolarized light. In this generalization, the state of polarization is represented by a coherency matrix [20–22]. The coherency matrix is defined as

$$\hat{\rho} = \begin{pmatrix} \langle E_x(t)E_x^*(t)\rangle & \langle E_y(t)E_x^*(t)\rangle \\ \langle E_x(t)E_y^*(t)\rangle & \langle E_y(t)E_y^*(t)\rangle \end{pmatrix}, \quad (3)$$

where $\langle\cdot\rangle$ represents the infinite time average. Physically, the trace of the coherency matrix is equal to the intensity of the wave, averaged over time. After passing through a component with the Jones matrix $\hat{J}$, the initial coherency matrix is transformed into $\hat{J}\hat{\rho}\hat{J}^\dagger$. The coherency matrix can also be given by the Stokes parameters $s_0$, $s_1$, $s_2$, and $s_3$ [23],

$$\hat{\rho} = \frac{1}{2}\begin{pmatrix} s_0 + s_1 & s_2 + is_3 \\ s_2 - is_3 & s_0 - s_1 \end{pmatrix}. \quad (4)$$

If we consider the normalized Stokes parameters $s_1/s_0$, $s_2/s_0$, and $s_3/s_0$ as coordinates in the three-dimensional Stokes space, then the Jones vectors represent the points on the Poincaré sphere, while the coherency matrices represent the points on and inside the Poincaré sphere.

## III. EXPERIMENTAL SETUP

Figure 1 shows the experimental setup for testing the LGI in Eq. (1). The output of a laser can be approximated by the fundamental transverse Gaussian mode ($TEM_{00}$). For such a mode with initial state $\hat{\rho}$, the time evolution is equivalent to its propagation along the optical axis [Fig. 1(a)]. The unitary time evolution $\hat{U}$ of $\hat{\rho}$ is generated by two half-wave plates (HWPs) as shown in Fig. 1(b), where $\theta_1$ and $\theta_2$ are the angles between the fast axes of the respective HWP and the horizontal axis. The time evolutions in each time step are chosen to be the same for simplicity. The correlation functions $C_{21}$, $C_{32}$, and $C_{31}$ are measured using the respective setups shown in Figs. 1(c)–1(e). In these setups, the measurement on the coherency matrix is realized via a polarizing beamsplitter with realizations $+1$ for horizontal polarization and $-1$ for vertical polarization. Under these realizations, the observable being measured is the Pauli matrix $\hat{\sigma}_z$, i.e., $\hat{Q}(t_i) = \hat{\sigma}_z$ for $i = 1, 2, 3$. The measured intensities $I_{ij}(m,n)$ at the intensity detectors represent the portion of light after measuring $m$ and $n$ at $t_i$ and $t_j$, respectively. The measured correlation functions are given by the intensity measurements in Figs. 1(c)–1(e) as

$$C_{ij} = \frac{\sum_{m,n\in\{-1,1\}} mnI_{ij}(m,n)}{\sum_{m,n\in\{-1,1\}} I_{ij}(m,n)}. \quad (5)$$

Let us denote the left-hand side of the LGI in Eq. (1) $\mathcal{K} \equiv C_{21} + C_{32} - C_{31}$. To theoretically calculate $\mathcal{K}$ within the coherency matrix representation, we use the following equations to obtain the theoretical values of the intensities,

$$I_{21}(m,n) = \mathrm{Tr}(\hat{P}(m)\hat{U}\hat{P}(n)\hat{\rho}\hat{P}(n)\hat{U}^\dagger\hat{P}(m)), \quad (6)$$

$$I_{32}(m,n) = \mathrm{Tr}(\hat{P}(m)\hat{U}\hat{P}(n)\hat{U}\hat{\rho}\hat{U}^\dagger\hat{P}(n)\hat{U}^\dagger\hat{P}(m)), \quad (7)$$

$$I_{31}(m,n) = \mathrm{Tr}(\hat{P}(m)\hat{U}\hat{U}\hat{P}(n)\hat{\rho}\hat{P}(n)\hat{U}^\dagger\hat{U}^\dagger\hat{P}(m)), \quad (8)$$

$$\hat{P}(m) = \frac{1+m}{2}\hat{P}_H + \frac{1-m}{2}\hat{P}_V, \quad (9)$$

where $\hat{U}$ represents the total Jones matrix of the two HWPs in Fig. 1(b), $\hat{P}_H$ and $\hat{P}_V$ are the respective Jones matrices of a horizontal and a vertical linear polarizer, and $\mathrm{Tr}(\cdot)$ denotes the trace of a matrix. The total Jones matrix of the two HWPs is determined by $\theta_1$ and $\theta_2$. The intensities in Eqs. (6)–(8) can be used to calculate the theoretical values of the correlation functions in Eq. (5). From the above equations, we can obtain the following expression for the theoretical value of $\mathcal{K}$ in terms of the HWP angles $\theta_1$ and $\theta_2$:

$$\mathcal{K} = 2\cos(4\theta_1 - 4\theta_2) - \cos(8\theta_1 - 8\theta_2). \quad (10)$$

Note that $\mathcal{K}$ is independent of the initial state $\hat{\rho}$ [16].

## IV. RESULTS

Figure 2 shows the experimentally measured and theoretically calculated values of $\mathcal{K}$ for different initial states. The diagonal polarization state is generated by rotating the polarization of the linearly polarized laser beam with an HWP. The partially polarized initial state is generated by combining two orthogonal polarizations [24]. For both initial states, $\theta_1$ is fixed at 0° and $\theta_2$ is varied from 0° to 180°. The measurements were repeated five times. Both plots in Fig. 2 show violations of the LGI in Eq. (1). The main source of experimental error is the laser intensity fluctuations during measurement, as 12 intensity measurements are needed to calculate one data point in Fig. 2. The detector noise is comparatively small and negligible (<1%). Some of the violations shown in Fig. 2 are higher than the theoretical maximum, which is mostly a consequence of the intensity fluctuations between measurements. The method used to generate the partially polarized polarization state introduces more fluctuations, which is reflected by the error bars in Fig. 2.

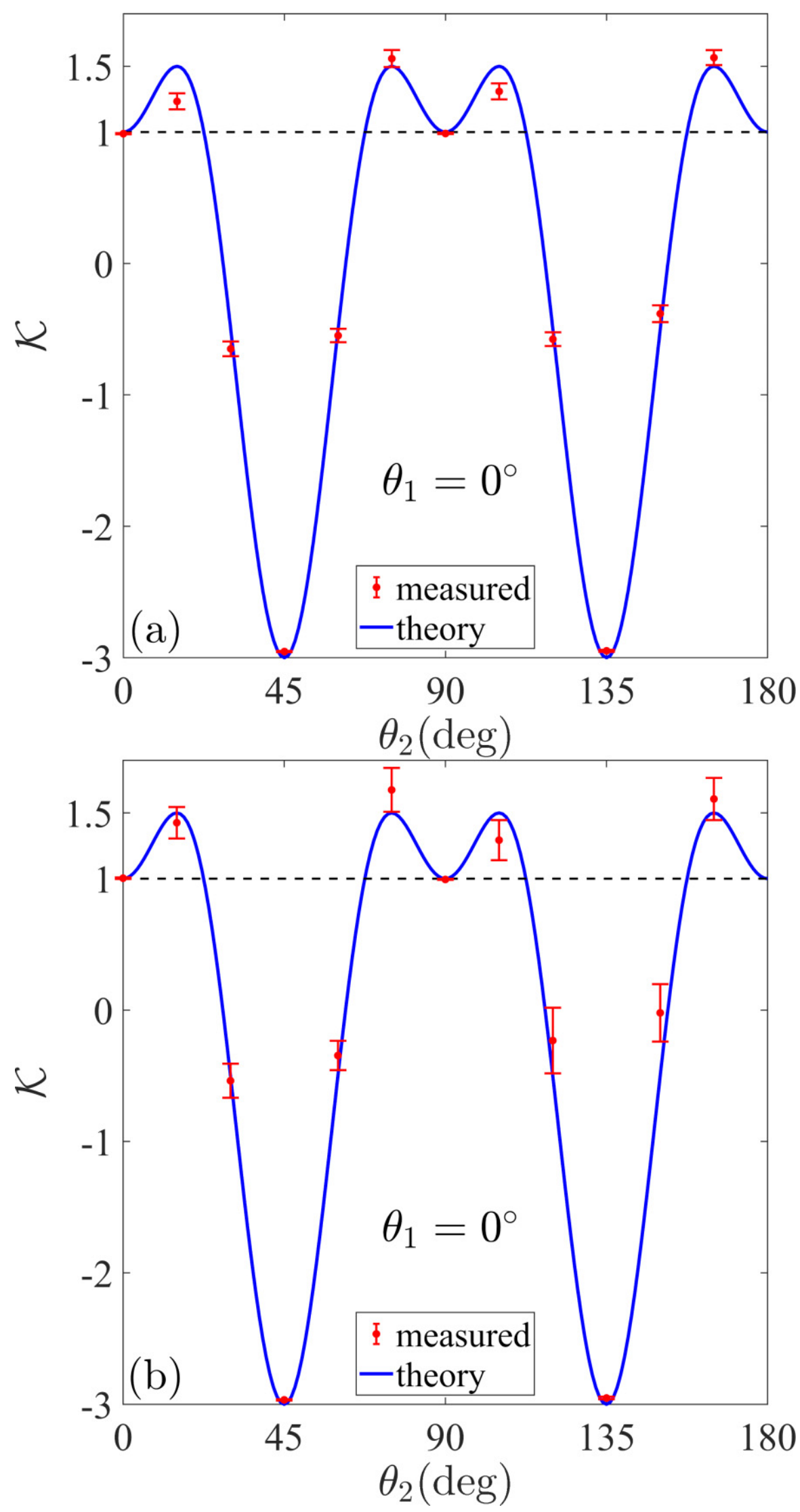

FIG. 2. Measured and theoretical values of $\mathcal{K}$ for initial states of (a) diagonal polarization and (b) a partially polarized state with degree of polarization ~0.8. In both plots, the data points above the dashed line are the violations of the LGI in Eq. (1). The theoretical values were calculated using Eq. (10).

## V. DISCUSSION

We have presented an experimental demonstration of violations of the Leggett-Garg inequality using only the polarization degree of freedom of a laser beam. Our results show that it is possible to violate the LGI in Eq. (1) using only classical linear polarization optics. The possibility of violating the LGI with the polarization of classical light is attributed to the superposition of temporally coherent fields [2,17]. It has been shown that the LGI can no longer be violated if the temporal degree of coherence is below a certain value [17]. However, as we did not quantify the invasiveness of our measurement apparatus, our results are still subject to the "clumsiness" loophole [1,25], which says that the violation of the LGI is not due to the fact that the system is nonmacrorealistic, but because the system was subjected to a measurement technique that happens to disturb it. This loophole can be closed by using ideal negative-result measurements [1] or adroit measurements [25]. For ideal negative-result measurements, we would need a device that interacts with the laser beam for only one particular state of polarization and does not interact at all otherwise. For adroit measurements, we would need to measure an additional observable, $\hat{\sigma}_\theta$, which can be realized by placing a series of retarders before the polarizing beamsplitter. However, even with these improvements, the stubborn macrorealist can still reject the violation results due to other weaker loopholes [8,25].

## ACKNOWLEDGMENTS

D.I.B. thanks Kurt Jacobs for insightful discussions. This work was supported by Defense Advanced Research Projects Agency (DARPA) Grant No. D19AP00043. The views and conclusions contained in this article are those of the authors and should not be interpreted as representing the official policies, either expressed or implied, of DARPA or the U.S. Government. The U.S. Government is authorized to reproduce and distribute reprints for Government purposes notwithstanding any copyright notation herein. R.T.G. acknowledges that this material is based upon work supported by, or in part by, the Army Research Laboratory and the Army Research Office under Contract/Grants No. W911NF-19-2-0087 and No. W911NF-20-2-0168. J.M.L. was supported by the Louisiana Board of Regents' Graduate Fellowship Program.